\documentclass[twocolumn,english,aps,pre,groupedaddress,showpacs]{revtex4-1}
\usepackage[T1]{fontenc}
\usepackage[latin9]{inputenc}
\usepackage{amsmath}
\usepackage{amssymb}
\usepackage{graphicx}
\usepackage{esint}

\makeatletter
\@ifundefined{textcolor}{}
{%
 \definecolor{BLACK}{gray}{0}
 \definecolor{WHITE}{gray}{1}
 \definecolor{RED}{rgb}{1,0,0}
 \definecolor{GREEN}{rgb}{0,1,0}
 \definecolor{BLUE}{rgb}{0,0,1}
 \definecolor{CYAN}{cmyk}{1,0,0,0}
 \definecolor{MAGENTA}{cmyk}{0,1,0,0}
 \definecolor{YELLOW}{cmyk}{0,0,1,0}
 }

\usepackage{dcolumn}
\usepackage{bm}
\usepackage{ulem}

\makeatother

\usepackage{babel}
\begin{document}

\title{Bistable collective behavior of polymers tethered in a nanopore}

\author{Dino Osmanovic}

\email[]{d.osmanovic@ucl.ac.uk}

\selectlanguage{english}%

\author{Joe Bailey}

\author{Anthony H. Harker}

\author{Ariberto Fassati}

\author{Bart W. Hoogenboom}

\email[]{b.hoogenboom@ucl.ac.uk}

\selectlanguage{english}%

\author{Ian J. Ford}

\email[]{i.ford@ucl.ac.uk}

\selectlanguage{english}%

\affiliation{London Centre for Nanotechnology, Department of Physics and Astronomy,
Centre for Mathematics and Physics in the Life Sciences and Experimental
Biology, and Wohl Virion Centre, Division of Infection and Immunity,\\
 University College London, Gower Street, London WC1E 6BT, United
Kingdom}

\date{\today}
\begin{abstract}
Polymer-coated pores play a crucial role in nucleo-cytoplasmic transport
and in a number of biomimetic and nanotechnological applications.
Here we present Monte Carlo and Density Functional Theory approaches
to identify different collective phases of end-grafted polymers in
a nanopore and to study their relative stability as a function of
intermolecular interactions. Over a range of system parameters that
is relevant for nuclear pore complexes, we observe two distinct phases:
one with the bulk of the polymers condensed at the wall of the pore,
and the other with the polymers condensed along its central axis.
The relative stability of these two phases depends on the interpolymer
interactions. The existence the two phases suggests a mechanism in
which marginal changes in these interactions, possibly induced by
nuclear transport receptors, cause the pore to transform between open
and closed configurations, which will influence transport through
the pore.
\end{abstract}


\maketitle

\section{Introduction}

Physical modeling is a powerful tool to interpret the complexity arising
from multiple interacting components in a biological system. One such
system is the nuclear pore complex (NPC). This structure mediates
all transport between the cell cytoplasm and the nucleus, and its
operation is thought to depend on the properties of natively unfolded
proteins, nucleoporins, that are end-grafted inside a $\sim$50~nm
wide channel \cite{Peters:2009,Wente:2010,Jamali:2011,Hoelz:2011}.
Molecules larger than $\sim$6~nm can only pass through this nanopore
if they are bound to nuclear transport receptors, which are known
to interact with the nucleoporins. Though the composition, hydrodynamic
size, and nanomechanical properties of single nucleoporins are known
in great detail \cite{Hoelz:2011,Yamada:2010,Lim:2006}, their collective
behavior in the NPC is still heavily disputed. This behavior has been
studied using a variety of models, with nucleoporins in a one-dimensional
geometry \cite{Zilman:2007}, grafted to a planar surface \cite{Miao:2009}
and constrained within a rectangular box \cite{Kustanovich:2004}.
Only very recently has the cylindrical geometry of the pore been taken
into account \cite{Peleg:2010,Moussavi:2011,Mincer:2011,Egorov:2011}.

More generally, there has been significant interest in polymer coatings
in nanopores, since they can be used to tune the aperture of artificial
and biomimetic nanopores and filters \cite{Iwata:1998,Wanunu:2007,Yameem:2009,Lim:2009,Jovanovic:2009,Kowalczyk:2011}.
Polymers end-grafted in a cylindrical pore have been studied by molecular
dynamics \cite{Adiga:2005} and Monte Carlo \cite{Koutsioubas:2009}
simulations. Depending on solvent quality, pore diameter and polymer
structure and dimensions, there can be a rich and complex phase diagram
\cite{Dimitrov:2006,Wang:2010,Peleg:2011}.

In this paper, we investigate the effect of confinement on the possible
conformations of polymers within a cylindrical channel. We have performed
Monte Carlo simulations on a coarse-grained model of polymers end-grafted
within a cylinder, and have furthermore developed an approach using
density functional theory (DFT) to study the relative stability of
competing morphologies. The DFT free energy is based on a reference
case of a tethered freely jointed chain of point-like beads, a novel
choice in this context, and its mean field variational optimization
has been carried out using a highly efficient numerical procedure.
The resulting well-founded free energy estimates enable us to construct
a phase diagram indicating the relative stability of different polymer
configurations and to speculate about transitions that might be induced
between them.

\begin{figure}
\includegraphics[width=1\columnwidth]{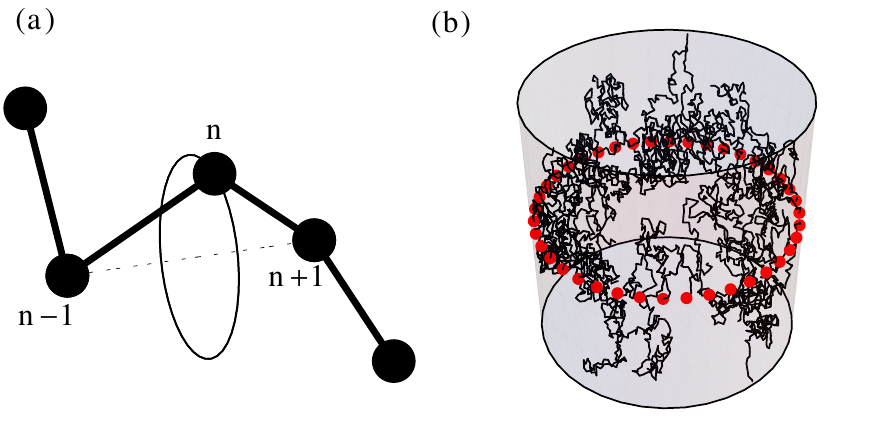}
 \caption{(Color online) (a) Illustration of the polymer model where bead $n$
is constrained to lie on a circle such that the length of the bonds
to beads $n-1$ and $n+1$ is unchanged. (b) Snapshot of a Monte Carlo
simulation with 40 non-interacting polymers of contour length 100~nm,
bead diameter $d=1$~nm and bond length $b=1$~nm, end-grafted to
the thick dots in a cylindrical pore of 25~nm radius. \label{fig:MC_schematics}}
\end{figure}

\section{Models of nucleoporin behavior}

Nucleoporins in the NPC channel have been shown to separate into two
distinct categories: those that form short globular conformations,
and longer polymers that tend to be able to extend further away from
their tethering point at the NPC rim \cite{Yamada:2010}. As cylindrical
confinement will mostly affect those nucleoporins with contour length
much greater than the pore radius, we focus on this latter category:
polymers with ~100~nm contour length, end-grafted on a ring around
the inner wall of a long cylinder with a radius $R=25$~nm (Fig.~\ref{fig:MC_schematics}).
We model the polymers as freely jointed chains of beads of diameter
$d=1$~nm, with a segment length $b=1$~nm. This implies a persistence
length of 0.5~nm, in agreement with single-molecule pulling experiments
on nucleoporin cNup153 \cite{Lim:2006}. Excluded-volume effects are
modeled as a hard-sphere repulsion between the beads. They are supplemented
by longer-ranged attractive interactions between the polymers, consistent
with those that appear to operate in the NPC \cite{Peters:2009,Wente:2010,Jamali:2011,Hoelz:2011}.
The combined bead-bead pair potential therefore takes the form
\begin{equation}
\phi(\mathbf{r})=\begin{cases}
\infty & |\mathbf{r}|<d\\
-\epsilon\exp[-(|\mathbf{r}|-d)/\lambda] & |\mathbf{r}|\geq d\,,
\end{cases}\label{eq:potential}
\end{equation}
 where $\mathbf{r}$ is the vector connecting the centers of the beads,
and $\lambda$ and $\epsilon$ are range and strength parameters,
respectively. A variety of interaction mechanisms might be represented
by an appropriate choice of $d$, $\lambda$ and $\epsilon$ in Eq.
~(\ref{eq:potential}), including, for example, the hydrophobic interaction
that is thought to play a significant role in these systems \cite{Israelachvili:2011}.

\begin{figure}
\includegraphics[width=1\columnwidth]{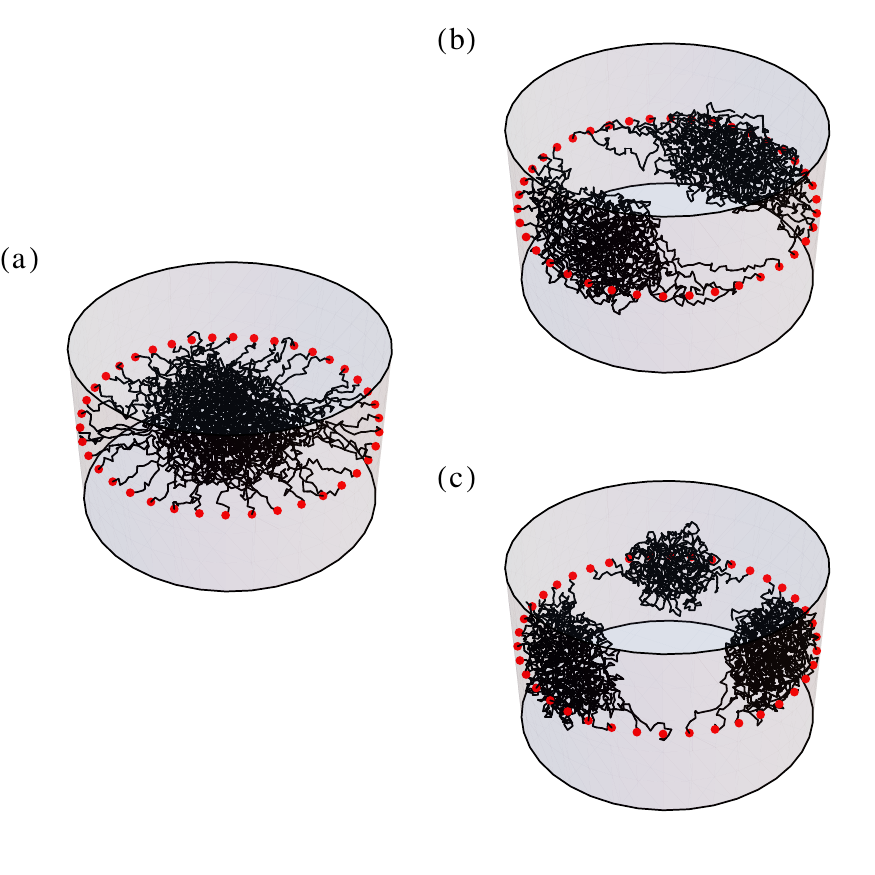}
\caption{(Color online) Snapshots taken from converged Monte Carlo simulations
as in Fig.~\ref{fig:MC_schematics}, showing different results for
polymers that are subject to the same excluded volume and attractive
interactions, as defined by Eq.~\ref{eq:potential} with $\epsilon=0.1\, k_{B}T$
and $\lambda=1.0$~nm. (a): Condensation in the center. (b) and (c):
Different numbers of clumps can be found when the polymers condense
closer to the wall. }

\label{fig:MC_snapshots}
\end{figure}

As a first approach to determine typical polymer configurations, we
study the system by Monte Carlo (MC) simulations. We employ a straightforward
Metropolis algorithm, in which single beads attempt moves on a circular
path defined by the constant distance $b$ to their nearest neighbors,
whilst remaining restricted to the inner volume of the cylinder as
illustrated in Fig.~\ref{fig:MC_schematics}. The first bead for
each polymer, located on the cylinder inner surface, is fixed in position,
whilst the last bead is free to move on the surface of the sphere
of radius $b$ centered on the penultimate bead. The restriction of
constant segment length simplifies the description of each MC move,
but can slow down the exploration of configuration space. Nevertheless,
all polymer conformations are accessible from one another. We perform
simulations of 250000 attempted moves per bead. Relaxation to equilibrium
is confirmed by noting convergence of the system energy, such that
we use the second half of each simulation to generate mean bead profiles.
The simulations indicate an interesting range of behavior, as can
be seen in Fig.~\ref{fig:MC_snapshots} for 40 polymers, each of
length 100 beads, tethered uniformly around a ring. Different dominant
configurations may be observed depending on the parameters chosen,
though they can be roughly divided into two categories: conformations
in which the density is peaked in the center, and those in which it
is peaked closer to the wall. Sometimes profiles from both categories
emerge even for the same parameter choice, depending on the starting
configuration. This suggests that there exist thermodynamically stable
and metastable states for a particular parameter set.

In order to categorize these phases more fully, and in particular
to investigate the relative stability of wall and central configurations,
a variational mean field density functional theory (DFT) of a many-polymer
system has been developed. DFT provides a natural framework for evaluating
the free energy of large numbers of interacting particles \cite{Evans:1979},
though certain modifications are necessary for it to be suitable as
a model of a system of polymers. It shares many of the features of
a self-consistent statistical field theory of polymers \cite{Frederickson:2006},
and has the capacity to include finite-range interactions, as expressed
in the pair-potential $\phi(\mathbf{r})$. The polymer entropy is
estimated by solving an equivalent Brownian motion problem, in contrast
to other approaches that are also based on the minimization of a free
energy functional but estimate the entropy by numerically generating
a large set of sample configurations \cite{Peleg:2011}. Our model
can be implemented using an efficient numerical algorithm that runs
on a standard desktop PC. It can be used to explore the equilibrium
behaviour of the system. Dynamical versions of DFT have been developed
to include relaxational phenomena \cite{Marconi:1999,Marconi:2007},
and our model has the potential to be extended in this direction.

The DFT model applied to a single chain of $N$ beads employs the
Hamiltonian
\begin{equation}
H=\sum_{i=0}^{N-1}h(\mathbf{r}_{i+1},\mathbf{r}_{i})+\frac{1}{2}\sum_{j=0}^{N}\sum_{i\ne j}^{N}\phi(\mathbf{r}_{i}-\mathbf{r}_{j})\,,\label{eq:2}
\end{equation}
 where $\mathbf{r}_{i}$ is the position of bead $i$, $h$ is a function
that constrains each segment of the polymer chain to take a fixed
length $b$, and the potential $\phi$ acts between all pairs of beads.
In the variational mean field approach, we introduce an additional
potential $V$ such that $H=H_{0}+H_{1}$ with
\begin{eqnarray}
H_{0} & = & \sum_{i=0}^{N-1}h(\mathbf{r}_{i+1},\mathbf{r}_{i})+\sum_{i=1}^{N}V(\mathbf{r}_{i})\label{eq:3}\\
H_{1} & = & \frac{1}{2}\sum_{j=0}^{N}\sum_{i\ne j}^{N}\phi(\mathbf{r}_{i}-\mathbf{r}_{j})-\sum_{i=1}^{N}V(\mathbf{r}_{i})\,.
\end{eqnarray}
 $H_{0}$ describes a chain of $N$ point-like beads interacting with
an external potential $V$, constrained by a requirement of constant
segment length, and with the first bead coupled to a static tethering
point at $\mathbf{r}_{0}$. This system has thermodynamic properties
embodied in the free energy \cite{Frederickson:2006}
\begin{equation}
F_{0}=-\ln\left(\int G(\mathbf{r}_{0},\mathbf{r},N;[w])\mathrm{d}\mathbf{r}\right)\,,\label{eq:4}
\end{equation}
 in units of $k_{B}T$, where $G(\mathbf{r}_{0},\mathbf{r},N;[w])$
is the propagator of a Brownian motion with contour length $Nb$ from
tether point $\mathbf{r_{0}}$ to point $\mathbf{r}$ inside the cylinder,
evolving under the influence of a dimensionless external potential
$w(\mathbf{r})=V(\mathbf{r})/k_{B}T$ acting as a sink; namely the
Green's function solution \cite{Doi:1986} to the diffusive problem
described by
\begin{equation}
\frac{\partial G(\mathbf{r}_{0},\mathbf{r},s;[w])}{\partial s}=\left(\frac{b^{2}}{6}\nabla^{2}-w(\mathbf{r})\right)G(\mathbf{r}_{0},\mathbf{r},s;[w])\label{eq:5a}
\end{equation}
 with initial condition $G=\delta(\mathbf{r}-\mathbf{r}_{0})$ at
$s=0$, and boundary conditions $G=0$ at the cylinder wall and zero
radial gradient at the center. $\rho(\mathbf{r})$ is the single bead
distribution function evaluated for the Hamiltonian $H_{0}$; it is
a functional of $w$, and is given in terms of $G$ as \cite{Doi:1986}
\begin{equation}
\!\rho(\mathbf{r})\!=\frac{\int_{0}^{N}\!\mathrm{d}s\int\!\mathrm{d}\mathbf{r^{\prime}}G(\mathbf{r}_{0},\mathbf{r},N-s;[w])G(\mathbf{r},\mathbf{r^{\prime}},s;[w])}{\int\mathrm{d}\mathbf{r^{\prime}}G(\mathbf{r}_{0},\mathbf{r^{\prime}},N;[w])}.\label{eq:7b}
\end{equation}

It is important to recognise that the mean field in DFT is a single-bead
potential that is introduced to emulate the real polymer self-interactions
as closely as possible. The reference Hamiltonian $H_{0}$ describes
the behavior of $N$ freely jointed beads in the potential $V$, and
the absence of additional bead-bead interactions simplifies its analysis
considerably. The DFT-derived bead density profiles represent polymer
configurations adopted in response to a mean field potential instead
of the actual self-interactions. This can be a reasonable approximation
if the mean field is optimized, as can be recognized through use of
the Bogoliubov inequality. Within such an approach, the best description
of the interacting polymer system can be obtained by minimizing the
free energy 
\begin{equation}
F_{{\rm mf}}\!=\! F_{0}-\!\int\!\!\rho(\mathbf{r})w(\mathbf{r})\mathrm{d}\mathbf{r}+F_{{\rm hc}}+\frac{1}{2}\!\int\!\!\rho(\mathbf{r})\rho(\mathbf{r^{\prime}})u(\mathbf{r}-\mathbf{r^{\prime}})\mathrm{d}\mathbf{r}\mathrm{d}\mathbf{r^{\prime}}\label{eq:7a}
\end{equation}
with respect to the mean field $w$, bearing in mind that $\rho$
depends on $w$. Both $F_{{\rm mf}}$ and $u$, the attractive part
of the bead pair potential $\phi$, are expressed in units of $k_{B}T$.
The contribution to the free energy arising from the attractive term
is derived on the basis of a random phase approximation. The contribution
from the repulsive interactions is modeled by a hard chain free energy
$F_{{\rm hc}}$ described in the local density approximation, and
in units of $k_{B}T$, by \cite{Wertheim:1987,Egorov:2008}:
\begin{align}
F_{{\rm hc}}= & \int\rho(\mathbf{r})\left[\frac{4\eta(\mathbf{r})-3\eta(\mathbf{r})^{2}}{[1-\eta(\mathbf{r})]^{2}}-\right.\label{eq:8a}\\
 & \left.\left(1-\frac{1}{N}\right)\ln\left(\frac{2-\eta(\mathbf{r})}{2[1-\eta(\mathbf{r})]^{3}}\right)\right]\,\mathrm{d}\mathbf{r}\nonumber
\end{align}
 where $\eta(\mathbf{r})$ is a bead packing fraction given by $\pi\rho(\mathbf{r})d^{3}/6$.
The origin of Eq. (\ref{eq:7a}) is described in detail in Appendix
A.

The model so far has been constructed for a single polymer, but a
system of $M$ polymers in a pore can be treated by multiplying $F_{0}$
by $M$, by interpreting $\rho$ in the remaining terms in Eq.~(\ref{eq:7a})
as the superposition of bead density profiles of the $M$ polymers
attached to their separate tether points, and by regarding the entire
free energy $F_{{\rm mf}}$ as a functional of a mean field $w$ that
we assume, for simplicity, to exhibit the cylindrical symmetry of
the pore.

\begin{figure}
\includegraphics[width=1\columnwidth]{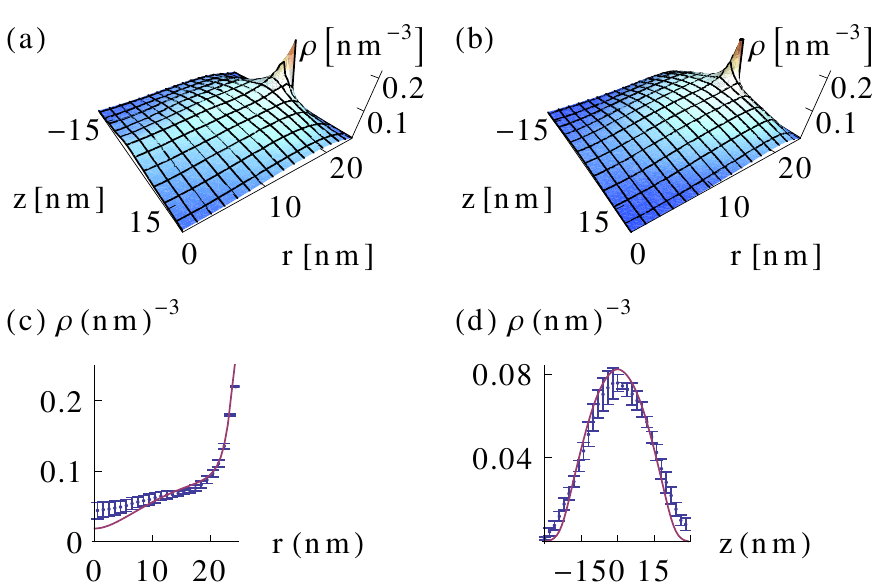}
\caption{(Color Online) Mean bead density profiles for Monte Carlo (MC) simulations
(a) and DFT calculations (b) for 40 polymers, composed of 100 beads
each, tethered within a cylinder, and interacting only through hard
sphere interactions. The plots are given for a range of radial ($r$)
and axial ($z$) coordinates. (c) and (d) show more detailed comparisons
of the profiles as a function of, respectively, the radial direction
in the plane of the tethering ring ($z=0$), and the axial direction
at a radius $r=17.5$~nm. DFT results are given by the solid lines
and the statistical uncertainty in the MC results is indicated with
error bars.}

\label{fig:MC_hs}
\end{figure}

We minimize the free energy functional (\ref{eq:7a}) using a conjugate
gradient method (Polak-Ribiere) \cite{Press:2007,Frederickson:2006},
on the basis of gradient information obtained by evaluating the functional
derivative $\delta F_{{\rm mf}}\left[\rho\right]/\delta\rho$. This
has proved to be more efficient in this context than employing the
gradient information within a steepest descent method: this is discussed
in greater detail in Appendix A. The numerical scheme involves making
an initial guess for $w(\mathbf{r})$, evaluating the bead density
using Eq. (\ref{eq:7b}), thereby obtaining the mean field free energy
through Eq. (\ref{eq:7a}) and using the conjugate gradient scheme
to generate a new free energy, and hence a new mean field. The mean
field and the bead density are defined numerically on a grid of points
within the cylinder, fine enough such that the spacing does not affect
the outcome of the calculations.

\section{Comparison between MC and DFT }

A key test of the DFT model, and of the numerical scheme, is to compare
bead density profiles with those obtained by MC simulation. We look
first at polymers interacting only through hard sphere repulsion,
represented in the DFT by the contribution in Eq. (\ref{eq:8a}).
The comparison shown in Fig. \ref{fig:MC_hs} shows a good agreement
between the two schemes, indicating that the DFT approach captures
the essence of the excluded volume behavior.

\begin{figure}
\includegraphics[width=1\columnwidth]{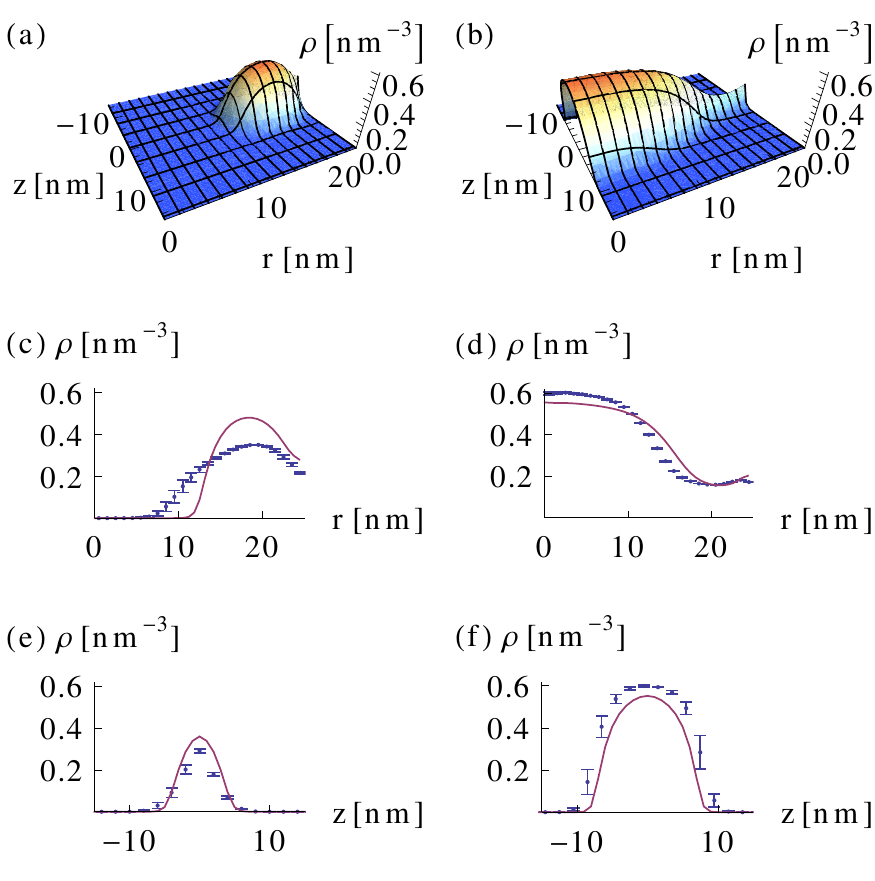}
 \caption{(Color online) Converged polymer configurations and bead density profiles
calculated for an attractive pair potential of depth $\epsilon=0.1\, k_{B}T$
and range $\lambda=1.0$~nm in a pore of 25~nm radius. The data
in the left and right column correspond to initial conditions with
the 40 polymers concentrated at the wall and at the center, respectively.
(a-b) DFT results with parameters corresponding to the Monte Carlo
simulations, as a function of radial ($r$) and axial ($z$) position.
(c-d) Comparison of Monte Carlo (with error bars) and DFT (smooth
curves) radial profiles for $z=0$, the plane of the tethering points.
Axial profiles (e) for the wall phase at $r=22.5$~nm and (f) for
the central phase at $r=2.5$~nm. \label{fig:MCDFTplots}}
\end{figure}

Next, we include long range interactions and make a similar comparison
between DFT and MC results. This test is more challenging since there
is now competition between polymer attraction and repulsion, giving
rise to quite distinct configurations, as we saw earlier for the MC
alone. Whilst the DFT can treat cases with different radial density
profiles, it presently does not explicitly allow for azimuthal clumping.
Fig.~\ref{fig:MCDFTplots}(a-b) illustrates two converged density
profiles obtained from the DFT approach, arising from different choices
for the initial mean field. They are the counterparts to the wall
and central phases observed in Monte Carlo simulations seen in Fig.
\ref{fig:MC_snapshots}, and use the same set of interaction parameters,
namely $\epsilon=0.1\, k_{B}T$ and $\lambda=1.0$~nm. A detailed
comparison of the centrally peaked profiles with MC results in Figs.~\ref{fig:MCDFTplots}(d)
and (f) suggests that the radial and axial spread of the polymers
determined from each approach are consistent with one another. For
the wall phase, the DFT and MC profiles do differ, as illustrated
radially and axially in Figs.~\ref{fig:MCDFTplots}(c) and (e). The
differences might be due to the angular symmetry breaking, or clumping,
observed in the Monte Carlo simulations, but this would not be expected
to affect the qualitative conclusions about the stability or metastability
of the two phases that we now explore.

The great benefit of the DFT model is that it provides thermodynamic
properties of the interacting polymers within the pore, not just the
structural properties that are available using MC. It provides estimates
of the free energy, such that it is possible to determine which phase,
central or wall, is thermodynamically stable or metastable under a
range of interaction conditions.

\begin{figure}
\includegraphics[width=1\columnwidth]{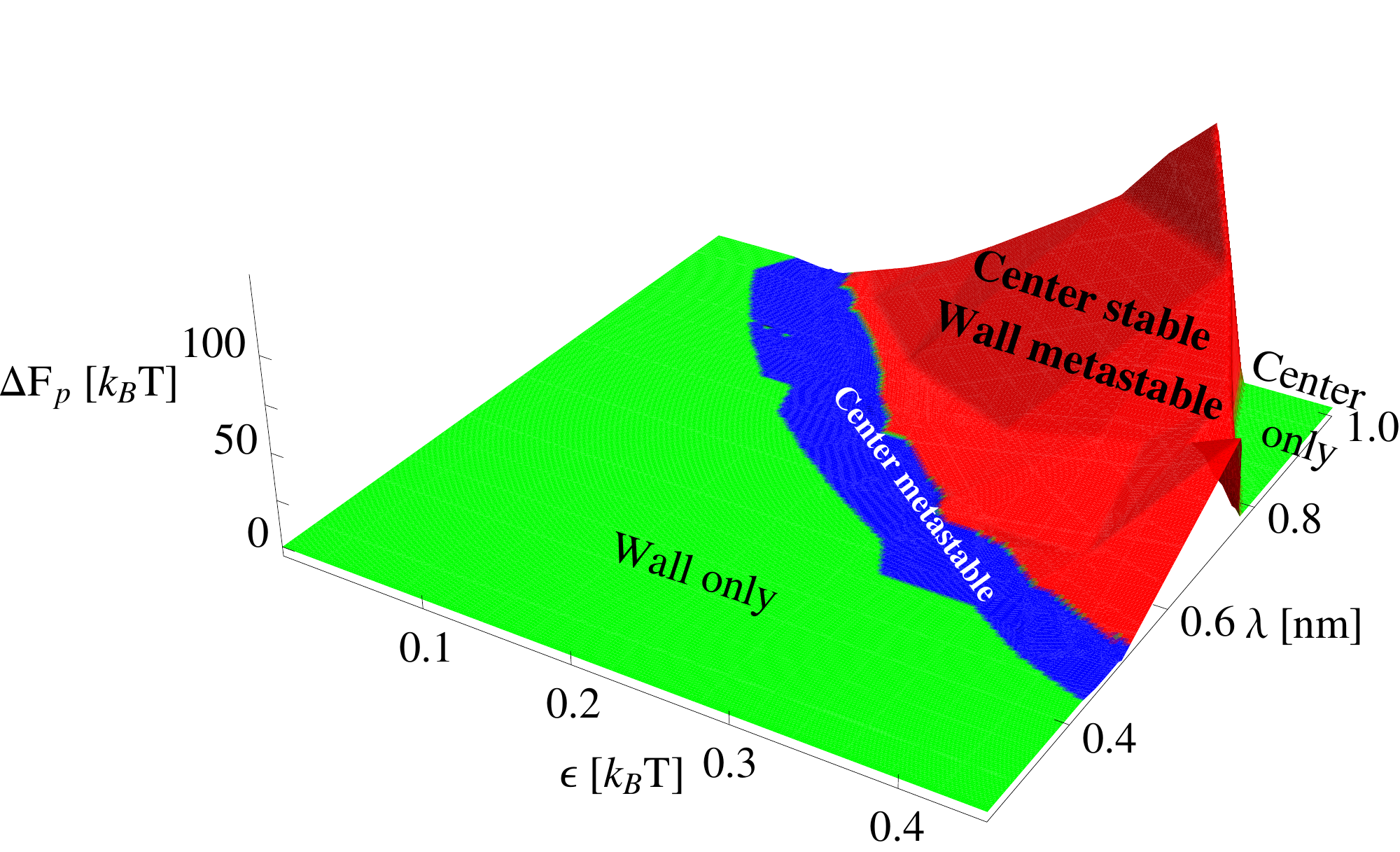}
 \caption{(Color online) Metastability diagram of the polymer phases as a function
of strength $\epsilon$ and range $\lambda$ of the attractive interactions.
$\Delta F_{\mathrm{p}}$ indicates the free energy difference per
polymer between the wall and central phases, for parameters where
they can both exist as a stable and metastable state. The coexistence
conditions lie along the boundary between blue and red regions. In
the green region, only one phase is found to be possible: the metastability
of the other has been lost. \label{fig:phasemap}}
\end{figure}

In Fig.~\ref{fig:phasemap} we plot the difference in free energy
between wall and central profiles $\Delta F_{\mathrm{p}}$, per polymer
and in units of $k_{B}T$, against interaction range $\lambda$ and
strength $\epsilon$. The plot illustrates the free energy difference
as a surface extending across regions where, respectively, the central
phase (`Center stable, wall metastable') and the wall phase (`Center
metastable') are thermodynamically stable and the other phase is metastable.
The region in the foreground (`Wall only') denotes conditions where
only the wall phase appears to exist. Similarly, there is a corresponding
region where only the central phase exists towards the back of the
diagram (`Center only'). There is a binodal line, or phase boundary,
where wall and central phase coexist, together with spinodals denoting
the extremes of metastability of one of the phases with respect to
the other. When we extend the range of $\epsilon$ to larger values,
i.e. stronger attractions than those shown, we find that the phase
boundary continues in such a manner such that a central phase is increasingly
favored.

The metastability diagram demonstrates that central polymer condensation
is a natural result of attractive interactions, provided that their
range and strength are sufficiently large. This is in accord with
intuition, which suggests that a central phase can be stabilized by
a reduction in energy to balance the cost in entropy of extending
the polymers away from the wall. A long range attractive potential
will favor this by allowing polymers to interact with more of their
counterparts across the other side of the pore. If the range were
reduced, then the required strength of the attraction would have to
be greater to produce the same effect. Repulsive interactions do not
drive bistable phase behavior. Instead the polymer condensate becomes
more homogeneous as the strength of the repulsion is increased: such
behavior is equivalent to that of a good solvent.

\begin{figure}
\includegraphics[width=1\columnwidth]{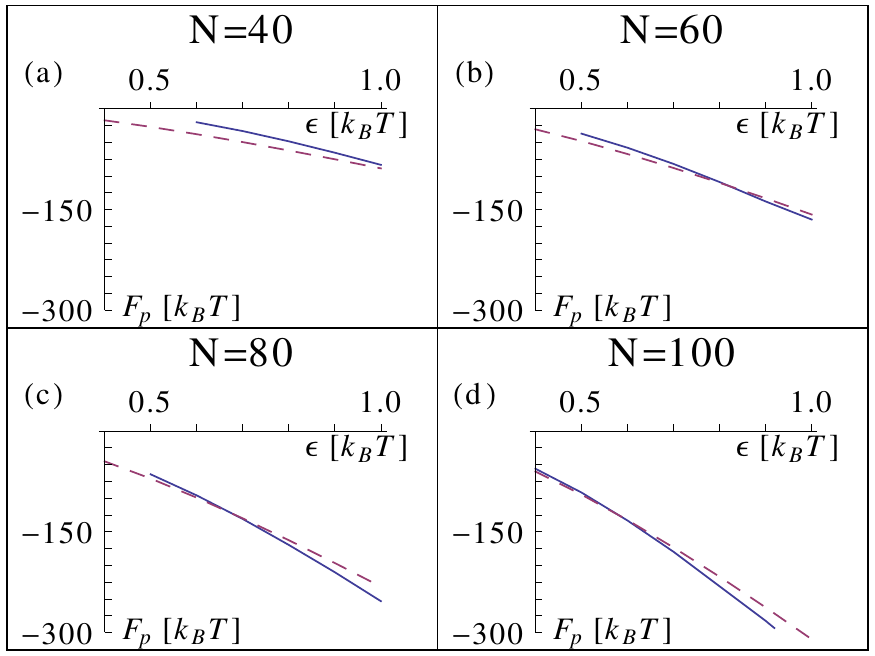}
 \caption{Free energies of the central phases (solid) and wall phases (dashed)
for systems with a range of bead number $N$, per polymer, for interaction
range $\lambda=0.5$ nm, plotted as a function of interaction strength
$\epsilon$. The phase with the lower free energy is thermodynamically
stable. The limited extent of the curve of higher free energy, in
some cases, is an illustration of the limits of the metastability
of that phase, equivalent to spinodal behavior. As $N$ increases,
the central phase becomes stable over a larger range of $\epsilon$.
\label{fig:phaselength}}
\end{figure}

In addition to changing the strength and range of the polymer interactions,
we can also explore the effect of a change in the number of beads
$N$ in each polymer. For constant segment length $b$, this is equivalent
to altering the length of the polymer. Reducing $N$ has the dual
effect of increasing the entropic cost of stretching a polymer towards
the center of the pore, and of decreasing the binding energy that
can be experienced by each polymer. Both effects hinder central phase
formation, and this is borne out in calculations of free energies
for the two phases as a function of interaction strength $\epsilon$
for a range of $N$, as shown in Fig.~\ref{fig:phaselength} for
$\lambda=0.5$ nm. Shortening the length of the polymers requires
a stronger attractive potential for the polymers to condense towards
the pore center, as is to be expected. \bigskip{}

\section{Discussion and Conclusions}

On the basis of Monte Carlo and density functional theory calculations,
we find that polymers tethered around a ring inside a cylindrical
geometry can exhibit bistable behavior, switching between a wall-
and a centrally condensed phase depending on the interaction parameters.
Interestingly, this behavior is observed for a geometry and polymer
structural properties that closely resemble the NPC \cite{Peters:2009,Wente:2010,Jamali:2011,Hoelz:2011},
for entirely realistic ranges ($\lambda\lesssim1$~nm) and strengths
($\epsilon\lesssim1\, k_{B}T$) of intermolecular interactions \cite{Israelachvili:2011}
within the NPC. The existence of a central polymer condensate is reminiscent
of the central `plug' or `transporter' structure of nucleoporins that
has been observed by cryo-electron microscopy of the nuclear pore
complex (NPC) \cite{Hoelz:2011,Yamada:2010}, though our model would
need to be developed to treat a more realistic geometry if it were
to be considered a proper representation. Nevertheless, our thermodynamic
model suggests that a phase transition from the central to the wall
phase occurs when the effective strength or range of the attraction
between polymer segments is decreased. Equivalently, the transition
might be induced by increasing the prevailing temperature. If we were
to apply our model to the NPC, this feature would be in agreement
with the experimentally observed dissolution of the central plug at
higher temperature \cite{Stoffler:2003}. A similar effect has been
seen upon incubation of the NPC with nuclear transport receptors \cite{Kramer:2011}.
A possible interpretation of this effect is that the nuclear transport
receptors are responsible for a weakening of the nucleoporin self-attraction
and that their influence on the polymer plug provides a mechanism
for the differential permeability of the NPC. In this scenario, a
sufficiently high concentration of nuclear transport receptors disturbs
the thermodynamic stability of the central structure to the extent
that the polymers withdraw towards the wall, leaving a free passage
to be occupied and traversed by receptors and receptor-bound cargos.
Such a mechanism, if it operates, could be exploited in the design
of artificial nanopores that might perform similar differential transport
of a variety of molecular species.


\begin{acknowledgments}
We gratefully acknowledge A. Kramer for discussions and T. Duke for
proofreading the manuscript. This work has been partially funded by
the Sackler Trust, the UK Biotechnology and Biological Sciences Research
Council (BB/G011729/1), the US Office for Naval Research (N00014-10­-1­-0096),
and the Wellcome Trust (083810/Z/07/Z).
\end{acknowledgments}
\appendix

\section{Details of the free energy functional}

We provide here some theoretical and numerical background, including
a detailed specification of the bead density profile, the rationale
for the variational principle employed, and the condition for minimizing
the free energy using functional derivatives. We focus our interest
on the statistical properties of a Hamiltonian $H_{0}$ for a freely
jointed polymer of $N$ point-like beads at positions $\{\mathbf{r}_{i}\}$
in a dimensionless external potential $w$, given by $H_{0}=U\left(\left\{ \mathbf{r}_{i}\right\} \right)+k_{B}T\sum_{i=1}^{N}w(\mathbf{r}_{i})$,
where $U\left(\left\{ \mathbf{r}_{i}\right\} \right)=\sum_{i}h(\mathbf{r}_{i+1},\mathbf{r}_{i})$
is a function that imposes the constraints on bond length. The 
bead at $\mathbf{r}_{1}$ is constrained with respect to the tether
point at $\mathbf{r}_{0}$. The partition function of the system is
\begin{equation}
Z_{0}=\int\prod_{j=1}^{N}\mathrm{d}\mathbf{r}_{j}\exp\left[-U/k_{B}T-\sum_{i=1}^{N}w(\mathbf{r}_{i})\right].\label{Deq:2}
\end{equation}
 The configuration-dependent bead density over continuous spatial
position $\mathbf{r}$ is defined as $\hat{\rho}(\mathbf{r},\left\{ \mathbf{r}_{k}\right\} )=\sum_{i=1}^{N}\delta(\mathbf{r}-\mathbf{r}_{i})$,
which allows us to write $Z_{0}$ as
\begin{equation}
Z_{0}[w]=\!\int[\mathrm{d}\mathbf{r}_{i}]\exp\left(-U/k_{B}T-\!\int\!\hat{\rho}(\mathbf{r})w(\mathbf{r})\mathrm{d}\mathbf{r}\right),\label{Deq:4}
\end{equation}
 where $[\mathrm{d}\mathbf{r}_{i}]$ represents the integration over
bead positions and we compress the notation of $\hat{\rho}$ for clarity.

The partition function is clearly a functional of the external potential
$w$ and its functional derivative is:
\begin{equation}
\begin{aligned} & \frac{\delta Z_{0}[w]}{\delta w(\mathbf{y})}=\\
 & \lim_{\epsilon\to0}\frac{1}{\epsilon}\!\left[\int\![\mathrm{d}\mathbf{r}_{i}]\exp\!\left(\!-\frac{U}{k_{B}T}-\!\int\!\!\hat{\rho}(\mathbf{r})\!\left(w(\mathbf{r})\!+\!\epsilon\delta(\mathbf{r}-\mathbf{y})\right)\!\mathrm{d}\mathbf{r}\!\right)\right.\\
 & \left.-\int[\mathrm{d}\mathbf{r}_{i}]\exp\left(-\frac{U}{k_{B}T}-\int\hat{\rho}(\mathbf{r})w(\mathbf{r})\,\mathrm{d}\mathbf{r}\right)\right]\\
 & =-\int[\mathrm{d}\mathbf{r}_{i}]\hat{\rho}(\mathbf{y})\exp\left(-U/k_{B}T-\int\hat{\rho}(\mathbf{r})w(\mathbf{r})\,\mathrm{d}\mathbf{r}\right),
\end{aligned}
\label{eq:5}
\end{equation}
 such that we can define $\rho(\mathbf{y})=\langle\hat{\rho}(\mathbf{y})\rangle$
to be the mean bead density for the system:
\begin{equation}
\begin{aligned}\rho(\mathbf{y}) & =\frac{1}{Z_{0}}\int[\mathrm{d}\mathbf{r}_{i}]\hat{\rho}(\mathbf{y})\exp\left(-U/k_{B}T-\int\hat{\rho}(\mathbf{r})w(\mathbf{r})\,\mathrm{d}\mathbf{r}\right)\\
 & =-\frac{1}{Z_{0}[w]}\frac{\delta Z_{0}[w]}{\delta w(\mathbf{y})}=-\frac{\delta\ln Z_{0}[w]}{\delta w(\mathbf{y})},
\end{aligned}
\label{DUPeq:6}
\end{equation}
 which demonstrates that $\rho$ is a functional of $w$. The explicit
functional dependence is given in Eq. (\ref{eq:7b}).

Now we discuss a self-interacting polymer described by the Hamiltonian
\begin{equation}
H=U+\frac{1}{2}\sum_{j=1}^{N}\sum_{i\ne j}^{N}\phi(\mathbf{r}_{i}-\mathbf{r}_{j})\,,\label{Deq:7}
\end{equation}
 incorporating a potential $\phi$ acting between all bead pairs.
We write $H=H_{0}+H_{1}$ with
\begin{equation}
H_{1}=\frac{1}{2}\sum_{j=1}^{N}\sum_{i\ne j}^{N}\phi(\mathbf{r}_{i}-\mathbf{r}_{j})-k_{B}T\sum_{i=1}^{N}w(\mathbf{r}_{i})\,,\label{Deq:8}
\end{equation}
 and employ the Bogoliubov inequality $F\le F_{0}+\langle H_{1}\rangle$,
where $F$ is the free energy of the system described by Hamiltonian
$H$, and $F_{0}=-\ln Z_{0}$ is the free energy of the reference
system described by $H_{0}$, in units of $k_{B}T$, and is given
by Eq. (\ref{eq:4}) \cite{Frederickson:2006}. As before, the brackets
denote an average over the ensemble associated with $H_{0}$, and
we write
\begin{equation}
\begin{aligned} & \left\langle \sum_{i=1}^{N}w(\mathbf{r}_{i})\right\rangle =\\
 & \frac{1}{Z_{0}}\int[\mathrm{d}\mathbf{r}_{i}]\!\int\!\hat{\rho}(\mathbf{y})w(\mathbf{y})\,\mathrm{d}\mathbf{y}\exp\left(-\frac{U}{k_{B}T}-\!\int\!\hat{\rho}(\mathbf{r})w(\mathbf{r})\,\mathrm{d}\mathbf{r}\!\right)\\
 & =\int\rho(\mathbf{y})w(\mathbf{y})\,\mathrm{d}\mathbf{y}.
\end{aligned}
\label{Deq:10}
\end{equation}
 In a similar fashion the mean of the pairwise terms is
\begin{equation}
\begin{aligned} & \!\!\left\langle \sum_{j=1}^{N}\sum_{i\ne j}^{N}\phi(\mathbf{r}_{i}-\mathbf{r}_{j})\right\rangle =\\
 & \!\frac{1}{Z_{0}}\!\int\![\mathrm{d}\mathbf{r}_{i}]\!\!\int\!\!\hat{\rho}_{2}(\mathbf{x},\mathbf{y},\left\{ \mathbf{r}_{k}\right\} )\phi(\mathbf{x}\!-\!\mathbf{y})\mathrm{\mathrm{d}\mathbf{x}d}\mathbf{y}\\
 & \!\!\times\!\exp\!\left(\!\!-\frac{U}{k_{B}T}-\!\!\int\!\!\hat{\rho}(\mathbf{r})w(\mathbf{r})\mathrm{d}\mathbf{r}\!\right)\!=\!\!\int\!\!\rho_{2}(\mathbf{x},\mathbf{y})\phi(\mathbf{x}-\mathbf{y})\mathrm{d}\mathbf{x}\mathrm{d}\mathbf{y},
\end{aligned}
\label{Deq:11}
\end{equation}
where $\hat{\rho}_{2}(\mathbf{x},\mathbf{y},\left\{ \mathbf{r}_{k}\right\} )=\sum_{i\ne j}\delta(\mathbf{x}-\mathbf{r}_{i})\delta(\mathbf{y}-\mathbf{r}_{j})$
is the two-point configuration-dependent bead distribution and $\rho_{2}(\mathbf{x},\mathbf{y})=\langle\hat{\rho}_{2}(\mathbf{x},\mathbf{y},\left\{ \mathbf{r}_{k}\right\} )\rangle$
is its mean in the $H_{0}$ ensemble. For simplicity, we take a random
phase approximation and represent $\rho_{2}(\mathbf{x},\mathbf{y})$
by $\rho(\mathbf{x})\rho(\mathbf{y})$. Thus the free energy of the
self-interacting polymer is bounded by the inequality
\begin{equation}
F\!\le\!-\ln Z_{0}[w]-\!\int\!\!\rho(\mathbf{y})w(\mathbf{y})\mathrm{d}\mathbf{y}+\frac{1}{2}\!\int\!\!\rho(\mathbf{x})\rho(\mathbf{y})\bar{\phi}(\mathbf{x}-\mathbf{y})\mathrm{d}\mathbf{x}\mathrm{d}\mathbf{y},\label{Deq:12}
\end{equation}
 where $\bar{\phi}$ is the pair potential divided by $k_{B}T$, which
can be separated into an attractive part $u(\mathbf{x}-\mathbf{y})$
and a repulsive part. The latter's contribution to the right hand
side may be represented by a functional $F_{{\rm hc}}[\rho]$ (Eq.
(\ref{eq:8a})) that has been found to capture the thermodynamic properties
of hard chains: freely jointed polymers of finite size hard spheres
\cite{Wertheim:1987,Egorov:2008}. The free energy then satisfies
\begin{equation}
\begin{aligned} & F\le F_{\mathrm{mf}}[w]=-\ln Z_{0}[w]+F_{{\rm hc}}[\rho]-\int\!\!\rho(\mathbf{y};[w])w(\mathbf{y})\,\mathrm{d}\mathbf{y}\\
 & +\frac{1}{2}\int\!\rho(\mathbf{x};[w])\rho(\mathbf{y};[w])u(\mathbf{x}-\mathbf{y})\,\mathrm{d}\mathbf{x}\mathrm{d}\mathbf{y},
\end{aligned}
\label{Deq:13}
\end{equation}
 where the functional dependence of $\rho$ on the mean field $w$
is explicitly noted.

The best estimate of $F$ is identified by functional minimization
of the mean field formulation $F_{\mathrm{mf}}[w]$ over all possible
$w$, to be achieved by setting the functional derivative $\delta F_{\mathrm{mf}}/\delta w(\mathbf{r})$
to zero. Several contributions to the derivative arise. We already
have $\delta\ln Z_{0}/\delta w(\mathbf{r})=-\rho(\mathbf{r};[w])$
from Eq. (\ref{DUPeq:6}), and furthermore, regarding $F_{{\rm hc}}$
as a functional of either $w$ or $\rho$,
\begin{equation}
\frac{\delta F_{{\rm hc}}[w]}{\delta w(\mathbf{r})}=\!\!\int\!\frac{\delta F_{{\rm hc}}[\rho]}{\delta\rho(\mathbf{y})}\frac{\delta\rho(\mathbf{y};[w])}{\delta w(\mathbf{r})}\mathrm{d}\mathbf{y}=\!\!\int\!\!\mu_{{\rm hc}}(\mathbf{y)}\frac{\delta\rho(\mathbf{y};[w])}{\delta w(\mathbf{r})}\mathrm{d}\mathbf{y},\label{Deq:13a}
\end{equation}
 where $\mu_{{\rm hc}}$ represents the functional derivative of $F_{{\rm hc}}$
with respect to $\rho$, together with
\begin{equation}
\frac{\delta}{\delta w(\mathbf{r})}\!\int\!\!\rho(\mathbf{y};[w])w(\mathbf{y})\mathrm{d}\mathbf{y}\!=\rho(\mathbf{r};[w])+\!\int\!\frac{\delta\rho(\mathbf{y};[w])}{\delta w(\mathbf{r})}w(\mathbf{y})\mathrm{d}\mathbf{y}\label{eq:14}
\end{equation}
 and
\begin{equation}
\begin{aligned} & \frac{\delta}{\delta w(\mathbf{r})}\int\rho(\mathbf{x};[w])\rho(\mathbf{y};[w])u(\mathbf{x}-\mathbf{y})\mathrm{d}\mathbf{x}\mathrm{d}\mathbf{y}\\
 & =2\int\frac{\delta\rho(\mathbf{y};[w])}{\delta w(\mathbf{r})}\rho(\mathbf{x};[w])u(\mathbf{x}-\mathbf{y})\mathrm{d}\mathbf{x}\mathrm{d}\mathbf{y},
\end{aligned}
\label{eq:15}
\end{equation}
 giving the minimization condition as:
\begin{equation}
\begin{alignedat}{1} & \frac{\delta F_{\mathrm{mf}}[w]}{\delta w(\mathbf{r})}=-\!\int\!\frac{\delta\rho(\mathbf{y};[w])}{\delta w(\mathbf{r})}w(\mathbf{y})\mathrm{d}\mathbf{y}\!+\!\!\int\!\!\mu_{{\rm hc}}(\mathbf{y)}\frac{\delta\rho(\mathbf{y};[w])}{\delta w(\mathbf{r})}\mathrm{d}\mathbf{y}\\
 & +\int\frac{\delta\rho(\mathbf{y};[w])}{\delta w(\mathbf{r})}\rho(\mathbf{x};[w])u(\mathbf{x}-\mathbf{y})\,\mathrm{d}\mathbf{x}\mathrm{d}\mathbf{y}=0.
\end{alignedat}
\label{eq:16-1}
\end{equation}
 Since
\begin{equation}
\frac{\delta F_{\mathrm{mf}}[w]}{\delta w(\mathbf{r})}=\int\frac{\delta F_{\mathrm{mf}}[\rho]}{\delta\rho(\mathbf{y})}\frac{\delta\rho(\mathbf{y};[w])}{\delta w(\mathbf{r})}\,\mathrm{d}\mathbf{y},\label{eq:16a}
\end{equation}
 this is equivalent to the condition
\begin{equation}
\frac{\delta F_{\mathrm{mf}}[\rho]}{\delta\rho(\mathbf{y})}=-w(\mathbf{y})+\mu_{{\rm hc}}(\mathbf{y)}+\!\int\!\rho(\mathbf{x})u(\mathbf{x}-\mathbf{y})\mathrm{d}\mathbf{x}=0,\label{eq:17}
\end{equation}
 which can be regarded as a requirement that the optimal mean field
acting on each bead is a suitable embodiment of the pairwise interactions.

We now discuss algorithms to determine the optimal mean field $w$
and associated density $\rho$ that satisfy this condition. A steepest
descent method could be employed such that $\rho$ is updated incrementally
and repeatedly in a direction down the local slope of the $F_{\mathrm{mf}}[\rho]$
surface. This can be regarded as an evolution of $\rho$ in a fictitious
time $t$ according to
\begin{equation}
\frac{\partial\rho(\mathbf{y},t)}{\partial t}=-\frac{\delta F_{\mathrm{mf}}[\rho]}{\delta\rho(\mathbf{y},t)},\label{eq:17a}
\end{equation}
until convergence at a time-independent density profile where the
left hand side vanishes. But the right hand side of this equation
requires $w$ as a functional of $\rho$, which is not readily available.
It is easier to determine $\rho$ for a given $w$, through Eq. (\ref{eq:7b}),
and we might therefore consider a scheme
\begin{equation}
\frac{\partial w(\mathbf{r},t)}{\partial t}=-\frac{\delta F_{\mathrm{mf}}[w]}{\delta w(\mathbf{r},t)},\label{eq:18a}
\end{equation}
 but the problem here is that the right hand side, given by Eq. (\ref{eq:16a}),
requires a specification of the functional derivative $\delta\rho(\mathbf{y};[w])/\delta w(\mathbf{r})$.
Instead, we employ the following arguments to formulate a third scheme.
From the definition of $\rho$, we can write
\begin{align}
 & \frac{\delta\rho(\mathbf{y};[w])}{\delta w(\mathbf{r})}=-\frac{\delta}{\delta w(\mathbf{r})}\left[\frac{1}{Z_{0}[w]}\frac{\delta Z_{0}[w]}{\delta w(\mathbf{y})}\right]\nonumber \\
 & =\frac{1}{\left(Z_{0}[w]\right)^{2}}\frac{\delta Z_{0}[w]}{\delta w(\mathbf{r})}\frac{\delta Z_{0}[w]}{\delta w(\mathbf{y})}-\frac{1}{Z_{0}[w]}\frac{\delta^{2}Z_{0}[w]}{\delta w(\mathbf{r})\delta w(\mathbf{y})},\label{eq:190}
\end{align}
and in view of Eqs. (\ref{eq:5}) and (\ref{DUPeq:6}) this becomes
\begin{equation}
\frac{\delta\rho(\mathbf{y};[w])}{\delta w(\mathbf{r})}=\rho(\mathbf{r})\rho(\mathbf{y})-\left\langle \hat{\rho}(\mathbf{r})\hat{\rho}(\mathbf{y})\right\rangle .\label{eq:191}
\end{equation}
Using the definitions of $\hat{\rho}$ and $\hat{\rho}_{2}$ this
gives
\begin{equation}
\frac{\delta\rho(\mathbf{y};[w])}{\delta w(\mathbf{r})}=\rho(\mathbf{r})\rho(\mathbf{y})-\rho_{2}(\mathbf{r},\mathbf{y})-\left\langle \!\sum_{i}\!\delta(\mathbf{r}-\mathbf{r}_{i})\delta(\mathbf{y}-\mathbf{r}_{i})\!\right\rangle .\label{eq:192}
\end{equation}
The random phase approximation that we have employed asserts that
$\rho_{2}(\mathbf{r},\mathbf{y})=\rho(\mathbf{r})\rho(\mathbf{y})$,
so we have
\begin{equation}
\frac{\delta\rho(\mathbf{y};[w])}{\delta w(\mathbf{r})}\approx-\left\langle \sum_{i}\delta(\mathbf{r}-\mathbf{r}_{i})\delta(\mathbf{y}-\mathbf{r}_{i})\right\rangle ,\label{eq:193}
\end{equation}
and we reach the important conclusion that, if viewed as a matrix,
to this level of approximation the off-diagonal elements of $\delta\rho(\mathbf{y};[w])/\delta w(\mathbf{r})$
are zero, and the diagonal elements are never positive.

Now we define a functional $\mathcal{F}[w]$ that satisfies
\begin{equation}
\frac{\delta\mathcal{F}[w]}{\delta w(\mathbf{r})}=-\frac{\delta F_{\mathrm{mf}}[\rho]}{\delta\rho(\mathbf{r})},\label{eq:18c}
\end{equation}
and compare the variational properties of $\mathcal{F}$ with those
of $F_{\mathrm{mf}}$. In view of Eq. (\ref{eq:16a}), the mean field
at a stationary point of $\mathcal{F}$, where $\delta\mathcal{F}[w]/\delta w(\mathbf{r})=0$,
will minimize $F_{\mathrm{mf}}$. Next we establish that an increment
in $w$ in the direction of steepest descent of the functional $\mathcal{F}$
gives rise to a decrease in the value of the functional $F_{{\rm mf}}$.
We show this by evaluating the analog in function space of the dot
product between two gradient vectors. We calculate, using Eqs. (\ref{eq:193})
and (\ref{eq:18c}), the projection of the functional derivative of
$F_{{\rm mf}}[w]$ along the direction in $w$ space of the functional
derivative of ${\cal F}[w]$, namely the integral:
\begin{align}
 & \!\int\frac{\delta\mathcal{F}[w]}{\delta w(\mathbf{r})}\frac{\delta F_{\mathrm{mf}}[w]}{\delta w(\mathbf{r})}\mathrm{d}\mathbf{r}=-\!\int\!\frac{\delta F_{\mathrm{mf}}[\rho]}{\delta\rho(\mathbf{r})}\frac{\delta F_{\mathrm{mf}}[\rho]}{\delta\rho(\mathbf{y})}\frac{\delta\rho(\mathbf{y})}{\delta w(\mathbf{r})}\mathrm{d}\mathbf{y}\mathrm{d}\mathbf{r}\nonumber \\
 & \approx\left\langle \sum_{i=1}^{N}\frac{\delta F_{\mathrm{mf}}[\rho]}{\delta\rho(\mathbf{r}_{i})}\frac{\delta F_{\mathrm{mf}}[\rho]}{\delta\rho(\mathbf{r}_{i})}\right\rangle ,\label{eq:20}
\end{align}
so that $\int[\delta\mathcal{F}[w]/\delta w(\mathbf{r})][\delta F_{\mathrm{mf}}[w]/\delta w(\mathbf{r})]\mathrm{d}\mathbf{r}\ge0$
to this level of approximation. The functional $\mathcal{F}$ has
a minimum at the mean field $w$ that minimizes $F_{{\rm mf}}$, and
an infinitesimal move along a path of steepest descent of the $\mathcal{F}$
surface also takes us downhill on the $F_{{\rm mf}}$ surface.

All this provides support for an algorithm for identifying the optimal
mean field and free energy based on solving the equation
\begin{equation}
\frac{\partial w(\mathbf{r},t)}{\partial t}=-\frac{\delta\mathcal{F}[w]}{\delta w(\mathbf{r},t)}=\frac{\delta F_{\mathrm{mf}}[\rho]}{\delta\rho(\mathbf{r},t)},\label{eq:18b}
\end{equation}
which is free of the above-mentioned problems of schemes (\ref{eq:17a})
and (\ref{eq:18a}).

Putting this scheme into practice, and using Eq. (\ref{eq:17}), we
discretize the spatial variable and the time and employ a forward
Euler numerical scheme with update rule
\begin{equation}
\begin{aligned} & w^{n+1}(\mathbf{y}_{k})=w^{n}(\mathbf{y}_{k})+\Delta t\Bigl[-w^{n}(\mathbf{y}_{k})+\mu_{{\rm hc}}^{n}(\mathbf{y}_{k})\\
 & +\sum_{j}\rho^{n}(\mathbf{x}_{j})u(\mathbf{x}_{j}-\mathbf{y}_{k})\Delta\mathbf{x}\Bigr],
\end{aligned}
\label{eq:19}
\end{equation}
 where $n$ labels discrete time and subscripts $j$ and $k$ label
discrete spatial points, and $\Delta\mathbf{x}$ is the volume element.
At each iteration starting with a mean field $w^{n}(\mathbf{y}_{k})$,
we use Eqs. (6) and (7) to generate the reference system bead density
profile $\rho^{n}(\mathbf{y}_{j})$ that is associated with this choice.
Through Eq. (\ref{eq:19}) with a given timestep $\Delta t$ this
gives us a new mean field $w^{n+1}(\mathbf{y}_{k})$ and the process
is repeated until the change in mean field falls below a chosen tolerance.
The converged field gives a minimized free energy $F_{\mathrm{mf}}$
which provides an upper limit to the actual free energy $F$ of the
self-interacting polymer system.

In fact, it proves to be more efficient to conduct the minimization
of $F_{{\rm mf}}[w]$ using a modified conjugate gradient scheme.
It is well known that such a scheme normally takes the form of repeated
line minimization of the functional along directions chosen in $w$
space that are selected according to the local gradient $\delta F_{\mathrm{mf}}/\delta w(\mathbf{r})$
and the direction of the preceding line search. We employ the Polak-Ribiere
version of this scheme but our modification is to select directions
based on the functional derivative $\delta\mathcal{F}/\delta w(\mathbf{r})$
instead. This is in the same spirit as the use of Eq. (\ref{eq:18b})
instead of Eq. (\ref{eq:18a}). The minimization of $F_{{\rm mf}}$
is performed numerically by stepping along the chosen direction in
a space spanned by the discrete $w^{n}(\mathbf{y}_{k})$ until we
encounter a change in sign of the difference $\Delta F_{{\rm mf}}$
with respect to the previous value. A new search direction is then
established and the process repeated. The scheme appears to be numerically
robust in practice, an indication that our consideration of the properties
of the functionals $F_{{\rm mf}}$ and $\mathcal{F}$ is sound.


%


\end{document}